\documentclass[aps,prl,reprint,showpacs,superscriptaddress,10pt]{revtex4-1}

\usepackage{graphicx}
\usepackage{color}
\usepackage{amsmath}
\usepackage{amsfonts}
\usepackage{bm}
\usepackage{enumerate}
\usepackage{array}
\usepackage{tensor}

\definecolor{orange}{rgb}{0,0,0}

\begin{document}

\title{Proposed Rabi-Kondo Correlated State in a Laser-Driven Semiconductor Quantum Dot}

\author{B. Sbierski}
\affiliation{Institute for Quantum Electronics, ETH Z\"urich, CH-8093 Z\"urich, Switzerland}
\affiliation{Dahlem Center for Complex Quantum Systems and Institut f\"ur Theoretische Physik, FU
Berlin, D-14195 Berlin, Germany}
\author{M. Hanl}
\affiliation{Arnold Sommerfeld Center for Theoretical Physics and Center for NanoScience, LMU 
M\"unchen, D-80333 M\"unchen, Germany}
\author{A. Weichselbaum}
\affiliation{Arnold Sommerfeld Center for Theoretical Physics and Center for NanoScience, LMU 
M\"unchen, D-80333 M\"unchen, Germany}
\author{H. E. T\"ureci}
\affiliation{Institute for Quantum Electronics, ETH Z\"urich, CH-8093 Z\"urich, Switzerland}
\affiliation{Department of Electrical Engineering, Princeton University, New Jersey 08544, USA}
\author{M. Goldstein}
\affiliation{Department of Physics, Yale University, 217 Prospect Street, New Haven, Connecticut, 06520, USA}
\author{L. I. Glazman}
\affiliation{Department of Physics, Yale University, 217 Prospect Street, New Haven, Connecticut, 06520, USA}
\author{J. von Delft}
\affiliation{Arnold Sommerfeld Center for Theoretical Physics and Center for NanoScience, LMU 
M\"unchen, D-80333 M\"unchen, Germany}
\author{A. \.Imamo\u glu}
\affiliation{Institute for Quantum Electronics, ETH Z\"urich, CH-8093 Z\"urich, Switzerland}

\begin{abstract}
Spin exchange between a single-electron charged quantum dot and
itinerant electrons leads to an emergence of Kondo correlations. When
the quantum dot is driven resonantly by weak laser light, the
resulting emission spectrum allows for a direct probe of these
correlations. In the opposite limit of vanishing exchange interaction
and strong laser drive, the quantum dot exhibits coherent oscillations
between the single-spin and optically excited states. Here, we show that
the interplay between strong exchange and non-perturbative laser
coupling leads to the formation of a new non-equilibrium
quantum-correlated state, characterized by the emergence of a
laser-induced secondary spin screening cloud, and examine the implications for the emission spectrum.
\end{abstract}

\maketitle

\begin{figure}[b]
    \begin{center}
        \includegraphics[width=\columnwidth]{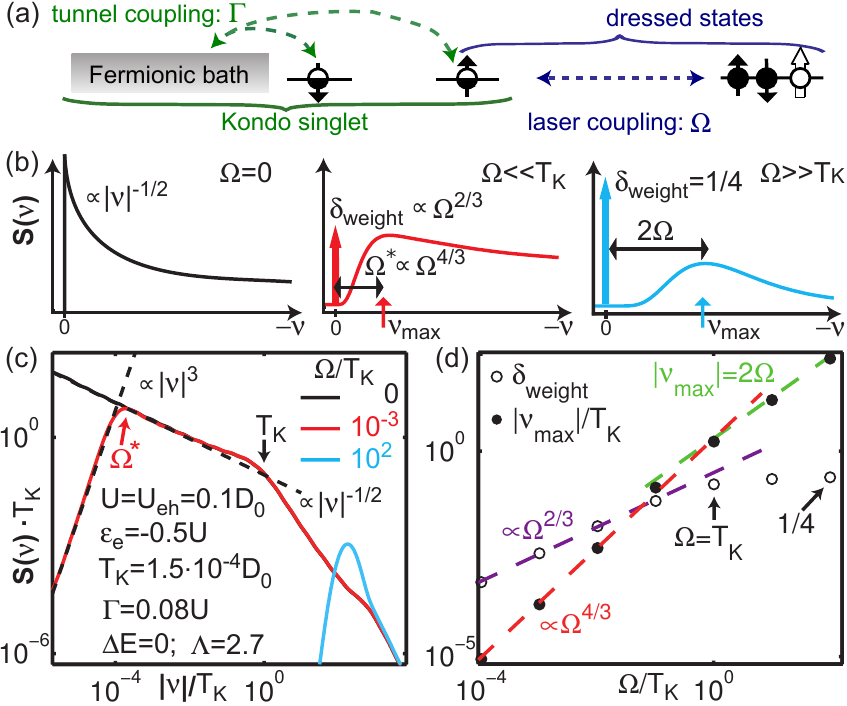}
        \caption{(Color online) (a) Competition between tunnel and laser coupling on the QD. While tunnel coupling favors a Kondo singlet state correlated with the FB, laser coupling favors dressed QD states. The characteristic energy scales are the Kondo temperature $T_\mathrm{K}$ and Rabi frequency $\Omega$, respectively. (b) Schematic plots of emission spectra
$S\left(\nu\right)$ from the Rabi-Kondo model Eq.~\eqref{eqn:hamiltonian} for $\Delta E=0$ and three characteristic
choices for $\Omega/T_{\mathrm{K}}$. For $\Omega = 0$, $S(\nu)$ shows a power-law divergence (left panel). With increasing laser intensity ($\Omega \neq 0$, other two panels) it transforms to a $\delta$-peak at $\nu = 0$ (thick arrows) and a broad maximum at the (renormalized) Rabi frequency.  (c), (d) 
NRG results with power-law asymptotes denoted as dashed lines: Log-log plot of the broad emission peak [$S\left(\nu<0\right)$] (c), its position $\left|\nu_{\mathrm{max}}\right|$ (dots) and the $\delta$-peak weight (circles)  vs.\ $\Omega/T_\mathrm{K}$ (d).
We have confirmed similar results
for the non-symmetric case $\varepsilon_{\mathrm{e}}\neq-U/2$ and
$U\neq U_{\mathrm{\mathrm{e}h}}$ as long as $n_{\mathrm{e}\uparrow}+n_{\mathrm{e}\downarrow}-n_{\mathrm{h}} \simeq1$.}
        \label{fig:RF results}
    \end{center}
\end{figure}

\emph{Introduction.---}
Exchange interactions between a singly-occupied quantum dot (QD) and a fermionic bath (FB) of itinerant electrons in the bulk lead to the formation of a Kondo state $\left|\mathrm{K}\right\rangle$ ~\cite{Hewson1993,Cox1998,Kouwenhoven2001}. When this many-body ground state is coupled by a laser field of vanishingly small Rabi frequency $\Omega$ to an optically excited trion state $\left|\mathrm{T}\right\rangle$ with an additional QD electron-hole pair [see Fig.~\ref{fig:RF results}(a)], the resulting emission spectrum at low FB temperatures $T$ is highly asymmetric~\cite{Tuereci2011,Latta2011}. Within the energy range defined by Kondo temperature $T_\mathrm{K} \gg T$, the spectral line shape is characterized by a power-law singularity. Anderson orthogonality (AO) determines the corresponding non-integer exponent and precludes any coherent light scattering in this limit. In the opposite limit of large $\Omega$ and vanishing exchange interaction ($T_K\to 0$), the emission spectrum consists of a Mollow-triplet and an additional $\delta$-function peak~\cite{Mollow1969,Flagg2009,Matthiesen2012}. While the latter stems from coherent Rayleigh scattering, the Mollow-triplet originates from incoherent transitions between dressed states which are superpositions of the original excited trion and the singly-charged ground states.

In this Letter, we analyze the interplay between strong exchange and non-perturbative laser couplings. By using a combination of numerical and analytical techniques, we find that the emission line-shape for $T \ll \Omega\ll T_K$ differs drastically from both the above limits. We demonstrate the emergence of a new quantum-correlated many-body state, which is a laser-induced, coherent superposition of the Kondo singlet state $\left|\mathrm{K}\right\rangle$ and the trionic state $\left|\mathrm{T}\right\rangle$ [see Fig.~\ref{fig:RF results}(a)]. The Kondo state involves a spin $1/2$ on the dot, screened by a spin cloud in the FB which is formed within distance $\propto 1/T_\mathrm{K}$ from the dot, while the FB is trivial in the bare trion state. The new quantum-correlated state is associated with the formation of an additional ``secondary" screening cloud at larger distances that compensates for the differences in local occupancies between $\left|\mathrm{K}\right\rangle$ and $\left|\mathrm{T}\right\rangle$. The secondary screening process is also of the Kondo type, and sets in below a secondary Kondo temperature, the renormalized Rabi frequency $\Omega^* \propto \Omega^{4/3}$. This new energy scale manifests itself in the location of a broad peak in the emission spectrum. The peak's red and blue tails follow power-law functions corresponding to the primary and secondary Kondo correlations, respectively. The emergence of the secondary screening cloud coincides with the recovery of the $\delta$-function peak in the emission spectrum, with weight scaling as $\Omega^{2/3}$. Measuring these effects should be possible in a setting similar to the one recently employed in Ref.~\cite{Latta2011}.  There the effects of Kondo correlations on the absorption spectrum of self-assembled QDs were measured in the limit $\Omega<T$, and the ability to resolve spectral features at $T<T_\mathrm{K}$ was demonstrated. Starting from this system one would need to increase the laser power to reach $\Omega > T$ while measuring the resulting resonance fluorescence spectrum; or, alternatively, to employ a continuous-wave laser pump-probe setup.

\emph{Model.---}
We consider a self-assembled QD in a semiconductor heterostructure, tunnel-coupled to a FB. We assume laser light propagating along the heterostructure
growth direction with right handed circular polarization $\sigma_{\mathrm{L}}=+1$
and a frequency $\omega_{\mathrm{L}}$ close to the QD trion
($\text{X}^{-}$) resonance. We model the system by an excitonic Anderson model \cite{Tuereci2011,SI}
augmented by a non-perturbative laser-QD interaction in the rotating wave
approximation. We set $\hbar=k_{\mathrm{B}}=1$ and assume zero magnetic field.
Optical selection rules imply that only the spin-down valence electron state will be optically excited, leading to the generation of a trion state involving a spin-up hole [Fig.~\ref{fig:RF results}(a)]. The spontaneous emission rate $\gamma_\mathrm{SE}$ is assumed to be negligibly small compared to all other energy scales. In the rotating frame, the Hamiltonian, to be called ``Rabi-Kondo model", thus reads:
\begin{align}
\label{eqn:hamiltonian}
 H = &
 \sum_{\sigma}\left(\varepsilon_{\mathrm{e}}-U_{\mathrm{eh}} \hat{n}_{\mathrm{h}}\right) \hat{n}_{\mathrm{e}\sigma}
 +U \hat{n}_{\mathrm{e}\uparrow}\hat{n}_{\mathrm{e}\downarrow}
 +\left(\varepsilon_{\mathrm{h}}-\omega_{\mathrm{L}}\right) \hat{n}_{\mathrm{h}}
 \nonumber \\ &
 + \sum_{k\sigma}\varepsilon_{k\sigma}c_{k\sigma}^{\dagger}c_{k\sigma}
 + \sqrt{\Gamma/\left(\pi\rho\right)} \sum_{k\sigma} \left( e_{\sigma}^{\dagger} c_{k\sigma} + \text{H.c.} \right)
 \nonumber \\ &
 + \Omega e_{\downarrow}^{\dagger}h^{\dagger}_\Uparrow+\text{H.c.}
\end{align}
The first line defines the QD Hamiltonian, 
where $\hat{n}_{\mathrm{e}\sigma}=e_{\sigma}^{\dagger}e_{\sigma}$, $\hat{n}_{\mathrm{h}}=h^{\dagger}_\Uparrow h_\Uparrow$,
while $e_{\sigma}^{\dagger}$ and $h^{\dagger}_\Uparrow$ are, respectively, creation operators
for QD spin-$\sigma$ electrons ($\sigma=\uparrow,\downarrow$ or $\pm 1$) and spin-up holes, $\varepsilon_e$ and $\varepsilon_h$ being the corresponding energies. We account for intra-dot Coulomb interaction by $U_{\mathrm{eh}}>0$
and $U>0$. To ensure a separated low-energy subspace formed by the states in Fig.~\ref{fig:RF results}(a), the
laser detuning from the bare QD transition, $\delta_{\mathrm{L}}=\omega_{\mathrm{L}}-\varepsilon_{\mathrm{e}}-U-\varepsilon_{\mathrm{h}}+2U_{\mathrm{eh}}$,
has to be small in the sense defined below.
The second line of Eq.~(\ref{eqn:hamiltonian}) models a noninteracting conduction band (the FB)  
of energies $\varepsilon_{k\sigma}\in[-D_0,D_0]$ with $\varepsilon_{\mathrm{F}}=0$ and constant density of
states $\rho \equiv 1/(2D_0)$ per spin, tunnel-coupled to the QD's $e$-level, giving it a width $\Gamma$. We assume $T\ll\Gamma\ll U\simeq U_{\mathrm{eh}}\ll D_0\ll\varepsilon_{\mathrm{h}},\,\omega_{\mathrm{L}}$
and investigate a situation where the QD carries one negative charge on average,
$n_{\mathrm{e}\uparrow}+n_{\mathrm{e}\downarrow}-n_{\mathrm{h}} \simeq1$ ~\cite{OneNegCharge}.
The QD-laser coupling of strength $\Omega$ [last term of Eq.~(\ref{eqn:hamiltonian})] connects the trion and Kondo subspaces, with projectors $P_\mathrm{T}=\hat{n}_h$ or $P_\mathrm{K}=1-\hat{n}_h$. When $\Omega=0$, these subspaces have hole and $e$-level occupancies $n_h^\mathrm{T}=1$ and $n^\mathrm{T}_{e \sigma}\simeq 1$ or $n_h^\mathrm{K}=0$ and $n^\mathrm{K}_{e \sigma}\simeq 1/2$, respectively, and ground states $\left|\mathrm{T}\right\rangle$ and $\left|\mathrm{K}\right\rangle$ with energy difference $\Delta E=E_{0\mathrm{,T}}-E_{\mathrm{0,\mathrm{K}}}$.

\emph{Emission spectrum.---}
The emission spectrum at detuning $\nu$ from
the laser frequency $\omega_{\mathrm{L}}$ is proportional to the
spectral function
\begin{equation}
S\left(\nu\right)=\sum_{n,m}\varrho_{m}\left|\left\langle n|h_\Uparrow e_{\downarrow}|m\right\rangle \right|^{2}\delta\left(E_{n}-E_{m}+\nu\right),\label{eq:S(nu)}
\end{equation}
where $\left|m\right\rangle $ and $E_{m}$ are eigenstates and eigenenergies
of the Rabi-Kondo model. We assume that spontaneous emission has a negligible effect
on the system's steady state, which is taken to be a thermal state
in the rotating frame at the temperature $T$ of the solid state environment, $\varrho_{m}=Z^{-1}e^{-E_{m}/T}$
\cite{SI}, and concentrate on $T=0$.
To simplify the discussion we will address mostly the $\Delta E=0$ case below (achieved by properly tuning the laser frequency $\omega_{L}$ to resonance), where the secondary screening effect is most pronounced, and defer the treatment of finite $\Delta E$ to the Supplemental Material (SM) \cite{SI}.
Fig.~\ref{fig:RF results}(b) schematically summarizes the main features of typical Numerical Renormalization Group (NRG) \cite{Bulla2008,Weichselbaum2007,Munder2012} results for the emission spectrum in Fig.~\ref{fig:RF results}(c,d).

For $\Omega \gg T_{\mathrm{K}}$ no signatures of Kondo physics are expected.
The emission spectrum can be completely
understood in terms of a dressed state ladder with the assumption
$\gamma_\mathrm{SE}=0$ and an intra-manifold, FB-induced
decay process \cite{Sbierski,DeltaPeakOmegaGrTK}.
The spectrum has two peaks: a broad peak at $\left|\nu_{\mathrm{max}}\right|=2\Omega$
and a delta-peak at $\nu=0$, both with equal weight $0.25$
[see Fig. \ref{fig:RF results}(b), blue line].

The situation is much more interesting for the Kondo-dominated regime, $\Omega \ll T_{\mathrm{K}}$, which we consider henceforth.
Here one might attempt to treat the QD-laser coupling [last term in Eq.~(\ref{eqn:hamiltonian})]
as a perturbation.
This would yield a spectrum that is essentially the same as the $\Omega=0$
spectrum calculated in Ref.~\onlinecite{Tuereci2011}.
However, we will show momentarily that this is correct only if the frequency $\left| \nu \right|$ is larger than a new energy scale $\Omega^* \ll T_\mathrm{K}$.

\emph{Effective model.---}
In order to understand this restriction on the perturbative treatment of the QD-laser coupling, as well as to derive the low-frequency behavior, we introduce an effective Hamiltonian $H^\prime$, which captures the essential physics of $H$ in the entire regime $|\nu|< T_{\mathrm{K}}$.
It can be thought of as the result of integrating out the degrees of freedom in the Rabi-Kondo model $H$ with energies larger than $T_\mathrm{K}$.
We can concentrate on just two states of the QD together with the surrounding FB degrees of freedom: The Kondo singlet state restricted (subscript r) to a FB region
of screening cloud size $\lesssim1/T_\mathrm{K}$, $\left|\mathrm{K}\right\rangle_\mathrm{r}$, and the trion state $\left| \text{T} \right\rangle_\mathrm{r}$, with no screening cloud. We thus replace the QD and the nearby degrees of freedom of the FB by a two level system (TLS) whose $\sigma^\prime_z = \pm 1$ ($\sigma^\prime_i$ being the Pauli matrices) eigenstates correspond to the $\left|\mathrm{T}\right\rangle_\mathrm{r}$ and $\left|\mathrm{K}\right\rangle_\mathrm{r}$, respectively. These are coupled by the laser and are split in energy. Furthermore, the outer electrons with energies $\lesssim T_\mathrm{K}$ experience different scattering phase shifts depending on the state of the TLS. Taking $\left| \text{K} \right\rangle_\mathrm{r}$  as reference state relative to which phase shifts
are measured, we have $\delta_\sigma^\text{K}=0$ and, by the Friedel sum rule \cite{Friedel}, $\delta_\sigma^{\text{T}}=\Delta_\sigma \pi$
where $\Delta_\sigma = (n^\mathrm{T}_{e \sigma}-\delta_{\sigma \downarrow})-n^\mathrm{K}_{e \sigma}=\sigma/2$ is the total dot charge difference per spin between $\left| \text{T} \right\rangle_\mathrm{r}$ and $\left| \text{K} \right\rangle_\mathrm{r}$. 
All this is captured by the following Hamiltonian:
\begin{equation}
H^\prime=
\sum_{k\sigma}\varepsilon^\prime_{k\sigma}c_{k\sigma}^{\prime \dagger}c^\prime_{k\sigma}
+
\Omega^\prime \sigma^\prime_{x}
+
\tfrac{\Delta E^\prime}{2}\sigma^\prime_{z}
+P_\mathrm{T}^\prime \sum_{\sigma,k,k^\prime}U^\prime_{\sigma} c^{\prime \dagger}_{k \sigma} c^\prime_{k^\prime \sigma}.
\label{eq:Hprime}
\end{equation} 
The first term describes the FB degrees of freedom whose distance from the QD is larger than $\sim1/T_K$, corresponding to a reduced half-bandwidth $D_0^\prime \sim T_\mathrm{K}$. The second term describes optical excitations, with $\Omega^\prime = \Omega \tensor*[_{\mathrm{r}}]{\left\langle \mathrm{K}|h_\Uparrow e_{\downarrow}|T\right\rangle}{_{\mathrm{r}}}\propto\Omega$.
The third term is the detuning, $\Delta E^\prime=\Delta E$ \cite{DeltaEprime}.
Finally, the last term accounts for the scattering of the FB electrons by the TLS, where $P_\mathrm{T}^\prime=(1+\sigma^\prime_{z})/2$ is a projector onto the trion sector. To reproduce the phase shifts mentioned above we choose
$U^\prime_{\sigma}$ equal to $- \sigma$ times a large positive numerical value ($\gg D_0^\prime$) which satisfies $\pi \rho^\prime U^\prime_\sigma=-\mathrm{tan}(\Delta_\sigma \pi)$.
NRG energy flow diagrams confirm that $H^\prime$ is a good description of the system below $T_\mathrm{K}$~\cite{SI}.
For $|\nu|<T_\mathrm{K}$, the emission spectrum $S(\nu)$ for the Rabi-Kondo model is reproduced qualitatively by $S^\prime \left(\nu \right)$ computed as in Eq.~\eqref{eq:S(nu)}, with $H^\prime$ and $\sigma^\prime_{-}$ replacing $H$ and $h_\Uparrow e_\downarrow$, respectively~\cite{SI}.

\emph{Intermediate-frequency behavior and emergence of a new energy scale.---}
To lowest order in $\Omega^\prime$, the behavior of $S^\prime(\nu)$ is governed by the AO between the TLS states $\left| \text{K} \right\rangle_{\mathrm{r}}$ and $\left| \text{T} \right\rangle_{\mathrm{r}}$, caused by the difference in phase shifts the FB electrons experience in the two states.
The spectrum thus behaves as a power law,
$S^\prime\left(\nu\right) \sim \left|\nu\right|^{2\eta^\prime_x-1}$,
with AO exponent
$2\eta^\prime_x =
  [\delta_\uparrow^\text{K}-\delta_\uparrow^{\text{T}}]^2/\pi^2 + [\delta_\downarrow^\text{K}-\delta_\downarrow^{\text{T}}]^2/\pi^2
  =1/2$ \cite{Nozieres1969,Munder2012}, in agreement with the $\Omega=0$ results of Ref.~\onlinecite{Tuereci2011}.
This implies that the hybridization operator $\sigma^\prime_x$ has a scaling dimension $\eta^\prime_x=1/4<1$ and is thus a relevant perturbation near the fixed-point $\Omega^\prime=0$.
Thus the leading-order renormalization group flow equation for $\Omega^\prime$ as one decreases the cutoff $D^\prime$ from its bare value $D^\prime_0$ is \cite{Cardy1996}
\begin{equation}
  D^\prime \tfrac{\text{d}}{\text{d} D^\prime} \left( \tfrac{\Omega^\prime}{D^\prime} \right)= \left( \eta^\prime_x-1 \right) \tfrac{\Omega^\prime}{D^\prime}.
\end{equation}
The dimensionless coupling $\Omega^\prime/D^\prime$ therefore grows and becomes of order one when the cutoff reaches the scale
\begin{equation} \label{eqn:omegas}
  \Omega^* = D^\prime_0 \left( \tfrac{\Omega^\prime}{D^\prime_0} \right)^{1/\left(1-\eta^\prime_x\right)} \sim T_{\mathrm{K}} \left(\tfrac{\Omega}{T_{\mathrm{K}}}\right)^{4/3} \ll T_{\mathrm{K}}.
\end{equation}
Hence, one may treat the term $\Omega^\prime \sigma^\prime_x $ [corresponding to the last term in Eq.~(\ref{eqn:hamiltonian})] as a perturbation
only if $\left|\nu\right| \gg \Omega^{*}$.
The power-law $S\left(\nu\right) \sim \left|\nu\right|^{-1/2}$ thus applies at intermediate frequencies, $\Omega^{*} \ll \left|\nu\right| \ll T_\mathrm{K}$.
The power-law divergence of the spectrum is cut off around $\left| \nu \right| \sim \Omega^*$ \cite{Munder2012}, resulting in a maximum in the spectrum at this scale, as confirmed by the NRG data shown in Fig.~\ref{fig:RF results}.
The emergence of this new energy scale is one of our central results.
At low frequencies, $\left|\nu\right| \ll \Omega^{*}$, the physics is governed by a new fixed point, which we now discuss.

\emph{Secondary Kondo screening.---} To understand this new fixed point we formally argue below that $H^\prime$ can be mapped onto the anisotropic Kondo model. This ``secondary'' Kondo model should not be confused with the original ``primary'' isotropic Kondo model for the QD spin. The role of the secondary Kondo temperature is played by $\Omega^*$; at energies below $\Omega^*$,
the original system flows to a strong-coupling fixed point featuring strong hybridization of Kondo and trion sectors, as confirmed by NRG level-flow data \cite{SI}. The low-energy behavior is universal when energies are measured in units of $\Omega^*$. 

\begin{figure}
    \begin{center}
        \includegraphics[width=\columnwidth]{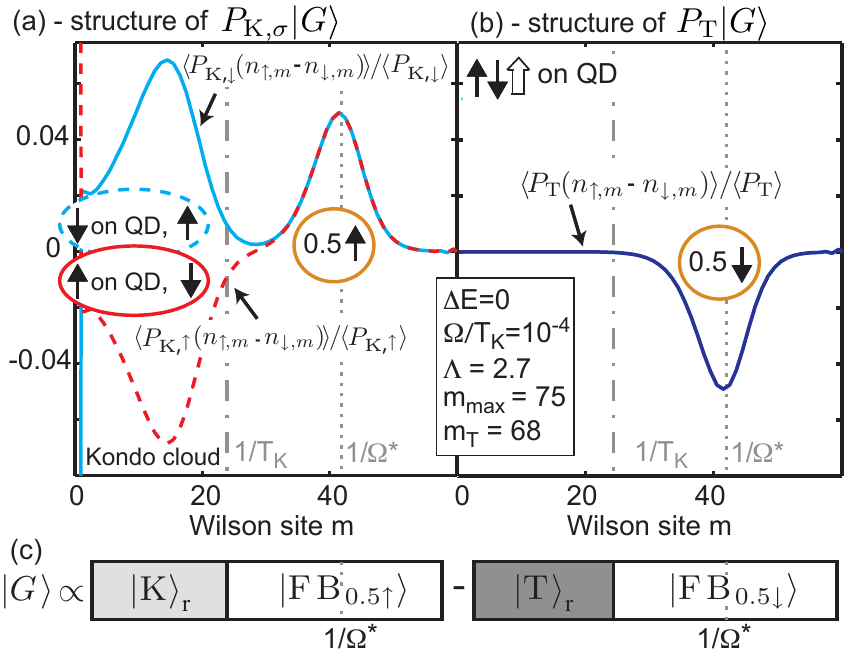}
        \caption{(Color online) NRG results for the Rabi-Kondo model Eq.~\eqref{eqn:hamiltonian}: Distribution of magnetization $n_{\uparrow,m}-n_{\downarrow,m}$ along the Wilson chain, obtained after acting with QD projectors (a) $P_{\mathrm{K},\sigma}=P_{\mathrm{K}}\hat{n}_{\mathrm{e},\sigma}\left(1-\hat{n}_{\mathrm{e},\bar{\sigma}}\right)$ 
and (b) $P_{\mathrm{T}}$, for $\Omega/T_{\mathrm{K}}=10^{-4}$
and $\Delta E=0$ at $T=2\cdot10^{-14} D_0$ (corresponding to the energy
scale of Wilson site $m_{\mathrm{T}}=68$). Even-odd oscillations have been averaged out. The QD is represented by site
zero. In (a), the Kondo state, restricted to a region
of size $\lesssim1/T_\mathrm{K}$ (subscript r), approximately  $\left|\mathrm{K}\right\rangle_\mathrm{r} \propto\left|\uparrow\right\rangle \left|\mathrm{FB_{\downarrow}}\right\rangle_\mathrm{r} -\left|\downarrow\right\rangle \left|\mathrm{FB_{\uparrow}}\right\rangle_\mathrm{r} $, 
features a singlet configuration
with FB magnetization opposite to that of the QD (left dip/peak structure). 
In (b)
the $\left|\mathrm{T}\right\rangle _{\mathrm{r}}$ trivial FB configuration appears within the same length scale.
The presence of a laser
($\Omega>0$) combines the states $\left|\mathrm{K}\right\rangle_\mathrm{r} $
and $\left|\mathrm{T}\right\rangle_\mathrm{r} $ into a new ground state shown in (c),
where FB states $\left|\mathrm{FB_{0.5\sigma}}\right\rangle $ carry
a magnetization of \textit{half} a spin $\sigma$ at distance $1/\Omega^{*}$
[peak in (a), dip in (b)]. The spin orientation of this secondary screening cloud is controlled by the laser polarization $\sigma_\mathrm{L}=+1$.}\label{fig:spin distribution: Omega>0}
    \end{center}
\end{figure}

One of the predictions of this ``secondary Kondo'' picture is that the ground state
of $H$ for $\Omega \ll T_K$ and $\Delta E=0$ is an equal-amplitude
superposition of the Kondo and trion states, with
some secondary screening cloud, whose distance from the QD is larger than the primary Kondo length $\propto 1/T_\mathrm{K}$.
To understand this nested screening cloud structure, consider $\left|\mathrm{K}\right\rangle $ (ground state for $\Omega=0$,
$\Delta E>0$) as a reference
state where the QD valence levels are filled and its conduction levels carry half an electron of each spin.
Since the total spin is zero, the correlation function between the QD spin and the total FB spin is $\left\langle S_{\mathrm{QD}}^{z}S_{\mathrm{FB}}^{z}\right\rangle =
\left\langle S_{\mathrm{QD}}^{z}\left(S_{\mathrm{FB}}^{z}+S_{\mathrm{QD}}^{z}\right)\right\rangle -\left\langle \left(S_{\mathrm{QD}}^{z}\right)^{2}\right\rangle
= -\left\langle \left(S_{\mathrm{QD}}^{z}\right)^{2}\right\rangle=
-1/4$.
This implies that when projecting into the subspace with spin up (down)
in the QD, the FB has a net additional single spin down (up)
electron \cite{Affleck} within a screening cloud up to a distance $\propto 1/T_\mathrm{K}$
from the QD [indicated by ellipses in Fig.~\ref{fig:spin distribution: Omega>0}(a)]. If, on the other hand, $\Omega=0$ but $\Delta E<0$,
the system is in the trion state $\left|\mathrm{T}\right\rangle$
with two QD electrons and a spin-up hole, and no screening cloud [Fig.~\ref{fig:spin distribution: Omega>0}(b)]. The absorbed $\sigma_\mathrm{L}=+1$ photon induces a change in QD charge per spin of $\Delta_\sigma$, thus causing the phase shifts $\delta^{\text{T}}_\sigma=\sigma\pi/2$ with respect to the reference
state, as mentioned above.
Turning on the laser source $\Omega$, when $\Delta E=0$,
the ground state is an equal-amplitudes superposition of the Kondo and trion states ($\left\langle P_{\mathrm{T}}\right\rangle =\left\langle P_{\mathrm{K}}\right\rangle =1/2$),
as depicted in Fig. \ref{fig:spin distribution: Omega>0}(c).
In analogy with the screening of a QD spin in a Kondo singlet, the FB screens the
spin configurations of the $\left|\mathrm{T}\right\rangle_\mathrm{r}$
and $\left|\mathrm{K}\right\rangle_\mathrm{r}$ states, which respectively have spin $\sigma/4$ or $-\sigma/4$ with respect to their mutual average of $\sigma/4$, by creating FB spin configurations with an opposite spin of $-\sigma/4$ or $\sigma/4$ (i.e. \textit{half} an electron spin) within distance $\sim 1/\Omega^*$ of the QD, respectively \cite{SzInteger}.
This nested screening cloud indeed appears in the NRG results in Fig.~\ref{fig:spin distribution: Omega>0}, further confirming our effective low-energy description.

\emph{Low-frequency behavior and $\delta$-peak.---}
To derive the low-frequency behavior of the spectrum, as well as the appearance of the elastic $\delta$-peak mentioned in the introduction, we make the notion of ``secondary Kondo effect'' more precise.
This can be done formally by transforming $H^\prime$ into a secondary Kondo model in two stages:
(i) Upon bosonization of the FB \cite{Gogolin1998} $H^\prime$ becomes the spin-boson model with Ohmic dissipation \cite{Weiss1999,Leggett1987}, the basic idea behind this mapping being that the low-lying particle-hole excitations of the FB are bosonic in nature, with a linear (Ohmic)
density of states;
(ii) the spin-boson model can be mapped onto the anisotropic Kondo model \cite{Hewson1993,Cox1998}:
\begin{align}
  H^\prime_{\mathrm{K}}
  = &
   -i v_F \sum_{\sigma=\uparrow,\downarrow} \int \text{d}x \psi_\sigma^\dagger(x) \partial_x \psi_\sigma (x) +
  \nonumber \\ &
  \tfrac{J_{z}}{2} S^\prime_{z} s^\prime_{z} \left(0\right) +
  \tfrac{J_{xy}}{2} S^\prime_{-} s^\prime_{+}\left(0\right) + \text{H.c.}
  -\Delta E_{z}S^\prime_{z}, \label{eq:H_K}
\end{align}
where $v_F=1/(\pi \rho^\prime)$ is the Fermi velocity, $S^\prime_i$ are the components of the secondary Kondo impurity spin, and $s^\prime_{i} \left(0\right)  = \sum_{\sigma,\sigma^\prime} \psi_\sigma(0)^\dagger \tau_i^{\sigma\sigma^\prime} \psi_{\sigma^\prime}(0) /2$ ($\tau_i$ being the Pauli matrices) are the FB spin density components in the vicinity of the impurity. Under this mapping
$\sigma^\prime_z = 2 S^\prime_z$ (hence $\Delta E_{z}\propto\Delta E^\prime$) but $\sigma^\prime_{+}\rightarrow S^\prime_{+}s^\prime_{-}\left(0\right)$.

One can now use known results on $H^\prime_{\mathrm{K}}$ to find the low-frequency ($\left| \nu \right| \ll \Omega^*$) behavior of the emission spectrum of $H^\prime$.
By the above mapping $S^\prime \left(\nu \right)$
is proportional to the spectral function of the retarded correlator
of $S^\prime_{+}s^\prime_{-}\left(0\right)$ with its conjugate in $H^\prime_{\mathrm{K}}$.
Since the anisotropic antiferromagnetic Kondo problem flows to the
same strong-coupling fixed point as the isotropic version, the calculation
of low-frequency power-law exponents can be done in the isotropic
case $J_{z}=J_{xy}$, where the $S^\prime_{+}s^\prime_{-}\left(0\right)$ correlator
can be replaced by the $S^\prime_{z}s^\prime_{z}\left(0\right)$ correlator.
At low energies, after the impurity spin is screened by the FB, $S^\prime_{z}s^\prime_{z}\left(0\right)$
can be replaced by the square of the local density of the $z$-component
of the electronic spin, which is a four-fermion operator. Thus, if the effective magnetic field vanishes, $\Delta E_z = 0$, its correlation function scales at
long times ($t > 1/\Omega^*$) as $t^{-4}$, leading to a $\sim\left|\nu\right|^{3}$
low-frequency behavior of the corresponding spectral function. The
same then applies to $S^{\left(\prime \right)}\left( \nu \right)$ in the regime
$\left|\nu\right| \ll \Omega^{*}$. This is indeed the behavior of the NRG results, cf.~Fig.~\ref{fig:RF results}(c) and \cite{SI}.

The above picture leads to another implication for the spectrum:
Since the relevant perturbation $\Omega^\prime \sigma_x^\prime$ strongly hybridizes, and thus cuts off the AO between the $\left| \text{K} \right\rangle$ and $\left| \text{T} \right\rangle$ states at energy scales smaller than $\Omega^*$ \cite{Munder2012}, a Dirac $\delta$-peak is now allowed to appear in the spectrum at $\nu=0$. By the definition of $S^\prime (\nu)$, its weight is $\delta_\text{weight} = \left| \left\langle \sigma_x^\prime \right\rangle \right|^2$.
Since $\Omega^*$ is the only low-energy scale, we expect that
$\Omega^\prime \left\langle \sigma_x^\prime \right\rangle \propto\Omega^{*}\propto\Omega^{\prime 4/3}$.
Hence, $\left\langle \sigma^\prime_x \right\rangle \propto\Omega^{\prime 1/3} \propto \Omega^{1/3}$, leading to
$\delta_{\mathrm{weight}}\propto\Omega^{2/3}$
which is in excellent agreement with the NRG results, Fig.~\ref{fig:RF results}(d) \cite{integralWeight,DeltaPeakKondo}.

\emph{Conclusions.---}
Laser excitation of a QD in the Kondo regime leads to a new correlated state featuring a nested spin screening cloud in the FB (Fig.~\ref{fig:spin distribution: Omega>0}) and gives rise to a specific double-peaked emission line shape (Fig.~\ref{fig:RF results}): (i) A broad peak centered at a renormalized Rabi frequency $\Omega^{*}$ [Eq.~\eqref{eqn:omegas}], with a $\sim\left|\nu\right|^{-1/2}$ red tail, resulting from the AO between ground and post-emission states resembling $\left|\text{T}\right\rangle$ and $\left|\text{K}\right\rangle$ at length scales $\ll 1/\Omega^*$. A Fermi-liquid type blue tail stems from the cut-off of the AO by the relevant Rabi coupling, as ground and post-emission state share a ``common'' FB configuration at length scales $\gg1/\Omega^*$ due to the secondary screening cloud. (ii) This common FB region leads to a $\delta$-peak at $\nu=0$ with weight $\propto\Omega^{2/3}$.  The $\Omega$-dependence of the coherent Rayleigh scattering strength in the presence of a finite spontaneous emission rate $\gamma_\mathrm{SE}$ remains an open question.

\emph{Acknowledgments.---}
We acknowledge helpful discussions with A. Rosch. This work was supported by an ERC Advanced Investigator Grant (B.S. \& A.I.). H.E.T. acknowledges support from the Swiss NSF under Grant No.~PP00P2-123519/1.
M.G.\ is supported by the Simons Foundation and the BIKURA (FIRST) program of the Israel Science Foundation,
L.I.G.\  by NSF DMR Grants 0906498 and 1206612, J.v.D., M.H. and A.W. by the DFG via NIM, SFB631, SFB-TR12, and WE4819/1-1.



\begin{thebibliography}{10}
\bibitem{DeltaEprime}
More precisely, an offset is added to $\Delta E^\prime$ such that $H^\prime|_{\Omega^\prime,\Delta E^\prime=0}$ has a ground state degenerate in the two TLS sectors.
\bibitem{SzInteger}
The total electron number is of course integer; if the screening cloud contains
  fractions of electrons, the rest is spread over the entire system, and thus
  does not contribute to the charge and spin distributions in the thermodynamic
  limit.
\bibitem{integralWeight}
The $\Omega$-scaling of the $\delta$-peak weight can also be found as follows: it scales as the spectral weight that goes missing from the $\Omega=0$ emission spectrum $\propto|\nu|^{-1/2}$ when the latter is cut off at $\nu \simeq \Omega^\ast$: $\int_0^{\Omega^{4/3}} d\nu\,\nu^{-1/2}\propto \Omega^{2/3}$.
\bibitem{DeltaPeakKondo}
However, further work including spontaneous emission in the theoretical
  treatment would have to revisit the actual form of the spectrum for
  $\left|\nu\right|<\gamma_\mathrm{SE}$ and the distribution of
  $\delta_{\mathrm{weight}}$ into an elastic and inelastic contribution.
\end{thebibliography}
\end{document}